\documentclass[10pt]{iopart}
\usepackage{graphicx,ifthen,iopams,setstack,amssymb,isolatin1}
\jl{1}
\newcommand{\writer}{michael}
\eqnobysec
\begin{document}
\title{Spinon excitations in the XX chain: spectra, transition rates, observability} 
\author{
Mitsuhiro Arikawa$^{1}$,  
Michael Karbach$^{2,3}$, 
Gerhard M{\"u}ller$^3$,
and Klaus Wiele$^{2,3}$
}
\address{
$^{1}$Yukawa Institute for Theoretical Physics, Kyoto University, 606-8502 Kyoto, Japan \\
$^2$Fachbereich Physik, Bergische Universit{\"a}t Wuppertal, 
  42097 Wuppertal, Germany \\
  $^3$Department of Physics,
  University of Rhode Island,
  Kingston RI 02881, USA \\
}

\ead{michael@karbach.org, gmuller@uri.edu}

\ifthenelse{\equal{\writer}{gerhard}}%
{\date{\today~--~1.0}} 
{\date{\version}}
\begin{abstract}
  The exact one-to-one mapping between (spinless) Jordan-Wigner lattice fermions
  and (spin-1/2) spinons is established for all eigenstates of the
  one-dimensional $s=1/2$ $XX$ model on a lattice with an even or odd number $N$
  of lattice sites and periodic boundary conditions. Exact product formulas for
  the transition rates derived via Bethe ansatz are used to calculate asymptotic
  expressions of the 2-spinon and 4-spinon parts (for large even $N$) as well as
  of the 1-spinon and 3-spinon parts (for large odd $N$) of the dynamic spin
  structure factors. The observability of these spectral contributions is
  assessed for finite $N$ and for $N\to\infty$.
\end{abstract}
%
\section{Introduction}\label{sec:intro}
%
The exact solution of the $XX$ model for exchange-coupled electron spins on a
one-dimensional lattice was first demonstrated more than 40 years ago
\cite{LSM61,Kats62}. Among all integrable quantum many-body systems, the $XX$
model is perhaps the one investigated most thoroughly
\cite{Niem67,KHS70,MBA71,BJ76,CP77,VT78,MS84a,MPS83a,MPS83b,IIKS93,SNM95,DKS00}.

The $XX$ interaction between nearest-neighbor spins with $s=1/2$ along
a chain of $N$ sites is described by the Hamiltonian
  \begin{equation}
    \label{eq:3}
    \mathcal{H}_{XX}=\sum_{\ell=1}^N[S_\ell^xS_{\ell+1}^x+ S_\ell^yS_{\ell+1}^y].
  \end{equation}
  The exact Jordan-Wigner mapping of this system onto a system of free, spinless
  lattice fermions has been at the root of most advances reported for this
  model. The many nontrivial properties of the $XX$ model are accounted for by
  the non-local functional relation between spin operators and fermion operators.
  
  In spite of steady progress on the $XX$ model reported since 1961, some
  questions of interest have resisted satisfactory answers.  The objective of
  this paper is to communicate significant advances on two such issues: (i) We
  establish an exact one-to-one mapping between the fermion composition and the
  spinon composition of all $XX$ eigenstates, thus linking the computationally
  convenient fermion quasiparticles to the physically relevant spinon
  quasiparticles on a rigorous basis. (ii) We use the exact product formulas
  for transition rates previously derived in the framework of the Bethe ansatz
  \cite{BKMW04} to calculate exact and asymptotic expressions of the $m$-spinon
  dynamic spin structure factors at $T=0$ with $m=2,4$ for even $N$ and $m=1,3$
  for odd $N$. The nature of the results raises questions regarding the
  observability of specific sets of $m$-spinon excitations and suggests that we
  distinguish a mesoscopic regime $(N\sim 10^2-10^3)$ from a macroscopic regime
  $(N\to\infty)$.

  The exclusion statistics of the fermion and spinon quasiparticles as well as
  the mapping between them is described in Sec.~\ref{sec:qpr}. The properties
  of $m$-spinon dynamic structure factors are discussed in
  Sec.~\ref{sec:dyspistruf} and the observability of $m$-spinon excitations in
  Sec.~\ref{sec:specwedi}.


%
\section{Fermions versus spinons}\label{sec:qpr}
%
It is desirable to carry out calculations for the $XX$ model using the fermions
because they are free and to interpret all spectral properties in terms of the
(interacting) spinons because the spinon vacuum, unlike the fermion vacuum, is
at the bottom of the spectrum.  Hence to calculate the $m$-spinon parts of the
$T=0$ dynamic spin structure factors from exact transition rate formulas based
on fermion momenta, we must understand in detail how the fermion and spinon
quasiparticles are related.

We consider the $XX$ model for even $N$ or odd $N$ and with periodic boundary
conditions. The Jordan-Wigner transform followed by a Fourier transform to the
reciprocal lattice converts Hamiltonian (\ref{eq:3}) into
\begin{equation}
  \label{eq:12a}
\mathcal{H}_{XX} = \sum_{\{k_i\}}\cos k_i\,c_{k_i}^\dagger c_{k_i}.
\end{equation}
Here $c_{k}^\dagger, c_{k}$ are fermion creation and annihilation operators. The sum
$\{k_i\}$ is over the allowed fermion momenta:
\begin{equation}
  \label{eq:13a}
 k_i=\frac{\pi}{N}\,\bar{m}_i,\quad
\bar{m}_i\in \left\{ 
\begin{tabular}{ll}
${\displaystyle \{1,3,\ldots,2N-1\}}$ & (even $N_f$) \\
${\displaystyle \{0,2,\ldots,2N-2\}}$ & (odd $N_f$)
\end{tabular} \right..
\end{equation}
The number of fermions varies over the range $0\leq N_f\leq N$.

%
\subsection{Exclusion statistics}\label{sec:exstat}
%
The number of eigenstates composed of $N_f$ fermions is
described by the combinatorial formula
\begin{equation}
  \label{eq:4a}
  W_f(N_f) = 
\left( \begin{tabular}{c} $d+N_f-1$ \\ $N_f$
  \end{tabular}\right),\quad d=N-N_f+1.
\end{equation}
The number of (spinless) fermions present
in an eigenstate is related to the quantum number $S_T^z$ ($z$-component of the
total spin) of that state as follows:
\begin{equation}
  \label{eq:15}
  N_f=\frac{N}{2}-S_T^z.
\end{equation}

The ground state of $\mathcal{H}_{XX}$ is non-degenerate for even $N$. It
contains $N_f=N/2$ fermions.  This state is reconfigured as the vacuum for
spinon quasiparticles. The (spin-1/2) spinons obey semionic exclusion
statistics. The number of eigenstates composed of $n_+$ spinons with spin up
and $n_-$ spinons with spin down is described by Haldane's generalization of
(\ref{eq:4a}) to fractional statistics \cite{Hald91a}:
\begin{equation}
  \label{eq:4b}
  W(n_+,n_-) = \prod_{\sigma=\pm}
\left( \begin{tabular}{c} $d_\sigma+n_\sigma-1$ \\ $n_\sigma$ \end{tabular}\right),
\end{equation}
\begin{equation}
  \label{eq:5b}
 d_\sigma =\frac{1}{2}(N+1) -\frac{1}{2}\sum_{\sigma'}(n_{\sigma'}-\delta_{\sigma\sigma'}).
\end{equation}
The total number of spinons, 
\begin{equation}
  \label{eq:36}
  n_++n_-=N_s, 
\end{equation}
is restricted to the values $N_s=0,2,\ldots,N$ for even $N$ and to the values
$N_s=1,3,\ldots,N$ for odd $N$.  Eigenstates with a fixed number of spinons may
contain different numbers of fermions and vice versa. However, from
(\ref{eq:15}) we infer
\begin{equation}
  \label{eq:27}
n_+ -n_-= N- 2N_f.
\end{equation}
Knowledge of $n_+$ and $n_-$ for a given eigenstate is thus equivalent to
knowledge of $N_s$ and $N_f$.

In Table~\ref{tab:WS6p7} we list the numbers of eigenstates
for $N=6$ (left) and $N=7$ (right) with given $N_s$ in the spinon
representation and given $N_f$ in the fermion representation.
\begin{table}[b]
  \caption{Number $W(n_+,n_-)$ of eigenstates with
    $N_s=n_+ +n_-$ spinons and (in a different representation)
    $N_f=N/2-(n_+ -n_-)/2$ fermions for $\mathcal{H}_{XX}$ with 
    $N=6$ (left) and $N=7$ (right). Subtotals (\ref{eq:lj25b})
    and (\ref{eq:lj25a}) are found in the last column and row,
    respectively.}\label{tab:WS6p7} \vspace*{3mm}
\centerline{
\begin{tabular}{c|rrrr|c}
$N_f\backslash N_s$ & 0 & 2 & 4 & 6 & \\ \hline
$0$~ & -- & -- & -- & 1 & 1 \\
$1$~ & -- & -- & 5 & 1 & 6 \\
$2$~ & -- & 6 & 8 & 1 & 15 \\
$3$~ & 1 & 9 & 9 & 1 & 20 \\
$4$~ & -- & 6 & 8 & 1 & 15 \\
$5$~ & -- & -- & 5 & 1 & 6 \\
$6$~ & -- & -- & -- & 1 & 1 \\ \hline
 & 1 & 21 & 35 & 7 & $W_f\backslash W_s$ \\ 
\end{tabular}\hspace{15mm}
\begin{tabular}{c|rrrr|c}
$N_f\backslash N_s$ & 1 & 3 & 5 & 7 & \\ \hline
$0$~ & -- & -- & -- & 1 & 1 \\
$1$~ & -- & -- & 6 & 1 & 7 \\
$2$~ & -- & 10 & 10 & 1 & 21 \\
$3$~ & 4 & 18 & 12 & 1 & 35 \\
$4$~ & 4 & 18 & 12 & 1 & 35 \\
$5$~ & -- & 10 & 10 & 1 & 21 \\
$6$~ & -- & -- & 6 & 1 & 7 \\
$7$~ & -- & -- & -- & 1 & 1 \\ \hline
 & 8 & 56 & 56 & 8 & $W_f\backslash W_s$ \\ 
\end{tabular}
}
\end{table} 
Summing the entries over $N_s$ yields the number of eigenstates with $N_f$
fermions, Eq.~(\ref{eq:4a}) rewritten as
\begin{equation}
  \label{eq:lj25b}
    W_f(N_f)= \left(
\begin{tabular}{c}
$N$ \\ $N_f$
\end{tabular}
\right).
\end{equation}
Summing the same entries over $N_f$ yields the number of eigenstates
with $N_s$ spinons,
\begin{equation}
  \label{eq:lj25a}
    W_s(N_s)= \left(
\begin{tabular}{c}
$N+1$ \\ $N_s$
\end{tabular}
\right),
\end{equation}
for a total number 
\begin{equation}
  \label{eq:28}
  \sum_{N_f}W_f(N_f) =\sum_{N_s}W_s(N_s)=2^N.
\end{equation}
The fermion vacuum is the entry on the upper right corner in both arrays. In
the array for $N=6$ the entry on the far left is the spinon vacuum, the
non-degenerate ground state for even $N$. The two entries on the far left in
the array for $N=7$ are 1-spinon states.  The fourfold degenerate ground state
for odd $N$ consists of two states from each entry.

%
\subsection{Spinon momenta}\label{sec:spimom}
%
The set of allowed spinon momentum values was first determined in the context of the
Haldane-Shastry (HS) model \cite{Hald91a} for even $N$. Here we adapt those
findings to the $XX$ model and extend them to lattices with odd $N$.  For this
purpose we only use symmetries common to the HS and $XX$ models: translations
along the (periodic) chain, rotations about the $z$-axis and the reflection $z \leftrightarrow
-z$ in spin space.

The momentum states available for occupation by spinons are equally spaced at
$\Delta\kappa=2\pi/N$ and their number is $N_0=(N-N_s)/2+1$. Hence the (generalized) Pauli
principle for spinons is semionic \cite{Hald91a}: $\Delta N_0=-g\Delta N_s$ with
$g=\frac{1}{2}$.
For even $N$ the allowed spinon momentum values are
\begin{equation}
  \label{eq:lj66}
  \kappa_i=\frac{\pi}{N}\,m_i,\quad m_i = \frac{N_s}{2}, \frac{N_s}{2}+2, \ldots,
  N-\frac{N_s}{2}.  
\end{equation}
Each available momentum state may be occupied by spinons of either spin
orientation without further restrictions. The quantum numbers $k$ and $S_T^z$
of all eigenstates are obtained from the spinon momenta $\kappa_j$ and
the spinon spins $\sigma_j$ via
\begin{equation}
  \label{eq:lj67}
  k=k_0+\sum_j\kappa_j,\quad S_T^z=\sum_j\sigma_j,
\end{equation}
where $k_0=0$ for even $N/2$ and $k_0=\pi$ for odd $N/2$.  For odd $N$ the
allowed spinon momentum values are
\begin{equation}
  \label{eq:lj68}
  \kappa_i=\frac{\pi}{N}m_i,\quad m_i= \frac{N}{2}+\frac{N_s}{2}, \frac{N}{2}+
  \frac{N_s}{2}+2, \ldots, \frac{3N}{2}-\frac{N_s}{2}. 
\end{equation}
The quantum numbers $k$ and $S_T^z$
are again determined by (\ref{eq:lj67})
but now with $k_0=0$ if $(N+N_s)/2$ is even and $k_0=\pi$ if
$(N+N_s)/2$ is odd.

%
\subsection{Fermion-spinon mapping}\label{sec:ferspim}
\begin{figure}[tb]
\centering
\includegraphics[width=48mm]{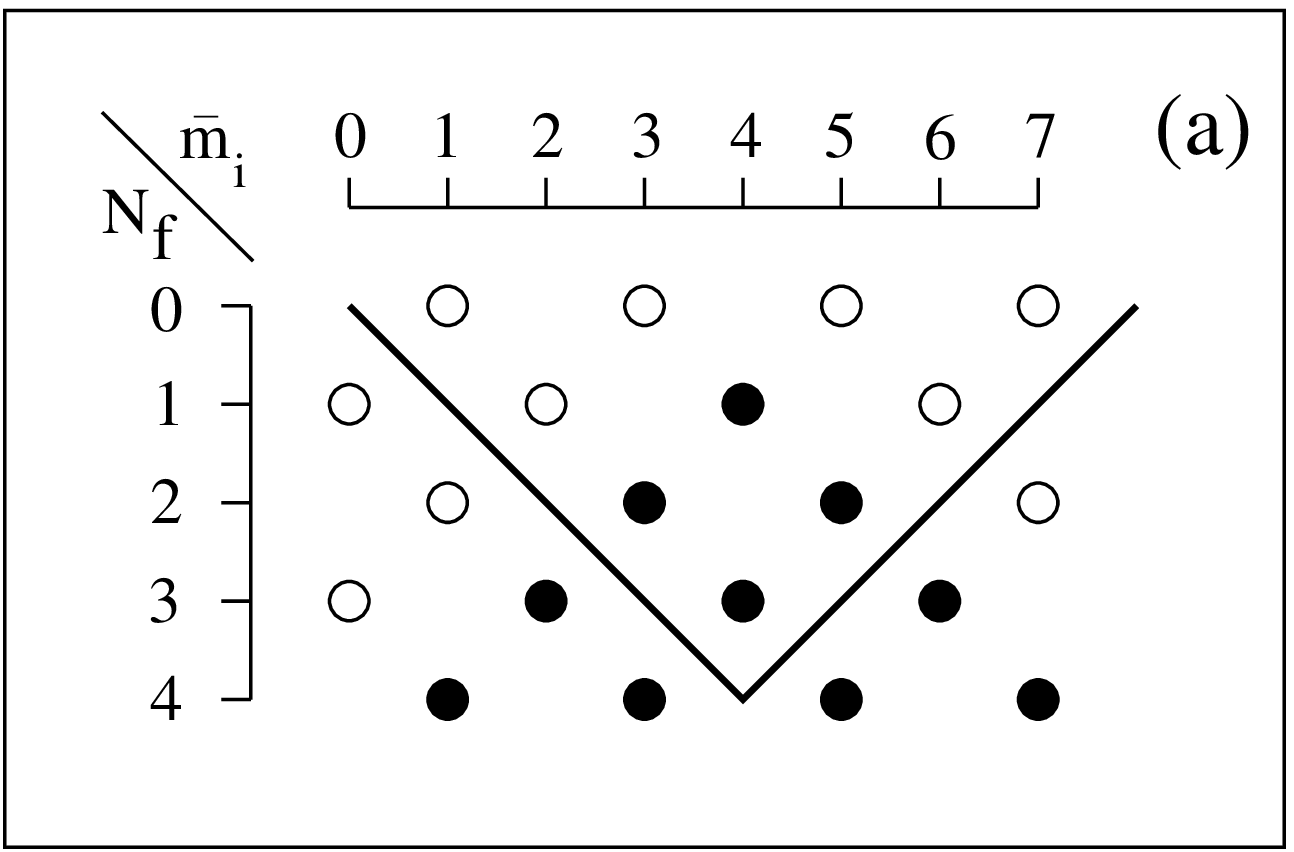}
\hspace*{5mm}%
\includegraphics[width=45mm]{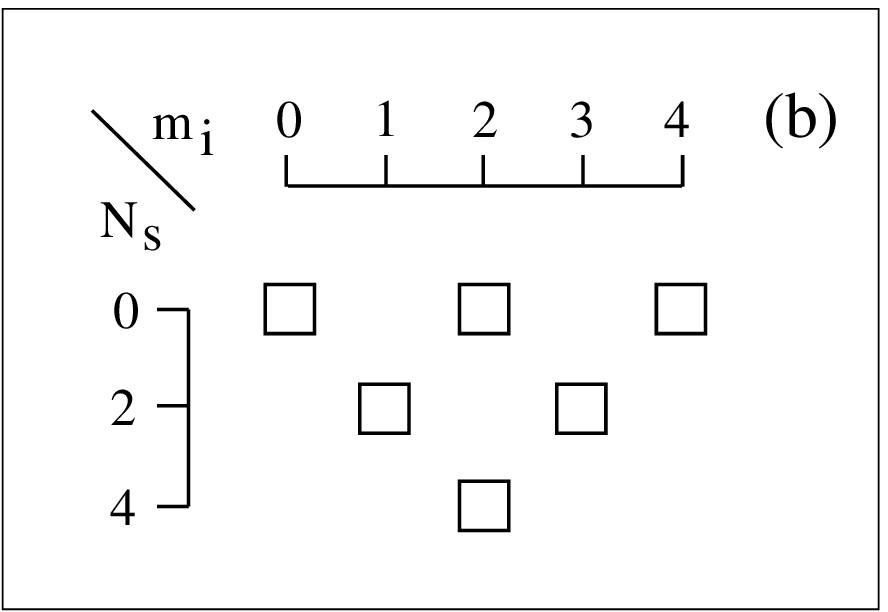}

\caption{(a) Fermion momentum states available to $N_f$ fermions and
  (b) spinon momentum states available to $N_s$ spinons in $XX$
  eigenstates for $N=4$. All momenta are in units of $\pi/N$. Fermion momentum
  states can be either vacant (open circle) or singly occupied (full circle).
  The particular fermion configuration shown in each row represents the lowest
  energy state for given $N_f$. Spinon momentum states can be either
  vacant or occupied by up to $N_s$ spinons with arbitrary spin polarization.
  No specific spinon configuration is shown.}
  \label{fig:fskN4}
\end{figure}

\begin{figure}[ht]
  \centering
  \includegraphics[width=75mm]{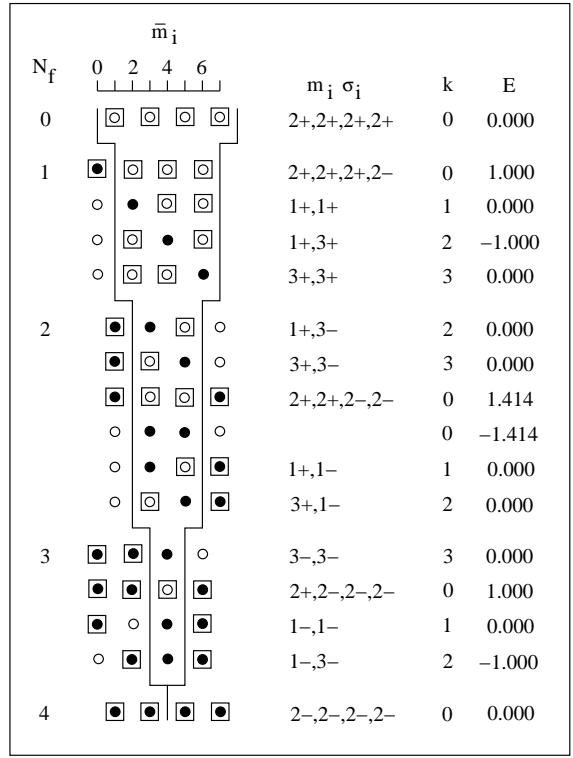}  
  \caption{Fermion configurations of all eigenstates for $N=4$. Fermionic
    particles (holes) are denoted by full (open) circles.  Spinon particles
    with spin up (down) are denoted by squares around open (full) circles. The
    fermion momenta $\bar{m}_i$ (in units of $\pi/N$) can be read off the
    diagram. The spinon momenta $m_i$ (also in units of $\pi/N$) and the spinon
    spins $\sigma_i$ are given explicitly and can be inferred from the fermion
    configuration as explained in the text. Also given are the wave number $k$
    (in units of $2\pi/N$) and the energy $E$ of each eigenstate.}
  \label{fig:fermspinN4}
\end{figure}

%
To keep track of the (physically relevant) spinons in the (computationally
convenient) fermion representation we consider a system of $N=4$ sites for
illustration. The allowed fermion momenta for $N_f=0,1,2,3,4$ are shown in
Fig.~\ref{fig:fskN4}(a) and the allowed spinon momenta for $N_s=0,2,4$ in
Fig.~\ref{fig:fskN4}(b).  An expanded version of Fig.~\ref{fig:fskN4}(a) is
shown in Fig.~\ref{fig:fermspinN4} with all $2^N=16$ distinct fermion
configurations with increasing $N_f$. The $\lor$-shaped line
of Fig.~\ref{fig:fskN4}(a) becomes the forked line in
Fig.~\ref{fig:fermspinN4}.

The exact spinon configuration is encoded in the fermion
configuration as described in the following:
(i) Consider the $\lor$ or the fork as dividing the fermion momentum
space into two domains, the inside and the outside. The outside domain wraps
around at the extremes ($\bar{m}_i=N\,\mathrm{mod}\,N=0$).
(ii) Every fermionic hole (open circle) inside represents a
spin-up spinon (square surrounding open circle) and every fermionic particle
(full circle) outside represents a spin-down spinon (square
surrounding full circle).
  (iii) Any number of adjacent spinons in the representation of
Fig.~\ref{fig:fermspinN4} are in the same momentum state of
Fig.~\ref{fig:fskN4}(b). Two spin-up (spin-down) spinons that are separated by
$\ell$ consecutive fermionic particles (holes) have momenta separated by
$\ell\pi/N$.
(iv) The spinon momenta are sorted in increasing order from the right-hand
  prong of the fork toward the left in the inside domain and toward the
  right with wrap-around through the outside domain.
  
  These rules determine the exact spinon spin and momentum configuration for the
  given fermion momentum configuration of all eigenstates pertaining to an $XX$
  chain with $N=4$. Generalizations to other even values of $N$ are
  straightforward. In applications of these rules to odd $N$ we must take note
  of the different set of allowed spinon momentum values and of the fact that
  the number of spinons is odd.

%
\section{Dynamic spin structure factors}\label{sec:dyspistruf}
%
The dynamic spin structure factors for a finite system (with even or odd $N$) can be
expressed as a weighted average
\begin{equation}
  \label{eq:29}
  S_{\mu\nu}(q,\omega) = \frac{1}{Z}\sum_ne^{-\beta E_n}S_{\mu\nu}(q,\omega)_n,\quad \mu\nu=+ -,- +,zz
\end{equation}
of functions
\begin{equation}
  \label{eq:lj36}
  S_{\mu\nu}(q,\omega)_n = 2\pi\sum_{n'} M_{nn'}^\mu(q)\delta\left(\omega-\omega_{nn'}\right),
\end{equation}
where $Z\doteq\sum_ne^{-\beta E_n}$ with $\beta\doteq(k_BT)^{-1}$ is the canonical partition function and
\begin{equation}
  \label{eq:trarat}
  M_{nn'}^\mu(q) =|\langle n|S^{\mu}_q|n'\rangle|^2
\end{equation} 
are transition rates for spin fluctuation operators
\begin{equation}
  \label{eq:30}
 S_q^\mu= \frac{1}{\sqrt{N}} \sum_{\ell=1}^Ne^{\imath q\ell}S_\ell^\mu
\end{equation}
with $q=(2\pi/N)m,~ m=0,1,\ldots,N-1$.

%
\subsection{Transition rates}\label{sec:prodexp}
%
Given the fermion momentum configurations
\begin{equation}
  \label{eq:31}
  \{k_1^n,\ldots,k_R^n\},\quad \{k_1^{n'},\ldots,k_r^{n'}\} 
\end{equation}
of the initial state $\langle n|$ and final state $|n'\rangle$ involved in
(\ref{eq:trarat}), the transition rates have the following explicit form
\begin{equation}
  \label{eq:lj37}
{M}^+_{nn'}(q) = 
  \frac{\left(\prod\limits_{i<j}^{R}\sin^{2}\frac{k_{i}^{n}-k_{j}^{n}}{2}\right)
\left(\prod\limits_{i<j}^{r}\sin^{2}\frac{k_{i}^{n'}-k_{j}^{n'}}{2}\right)}%
  {\prod\limits_{i=1}^{R}N^{2}\prod\limits_{j=1}^{r}\sin^{2}\frac{k_{i}^{n}-k_{j}^{n'}}{2}}\,\delta_{r,R+1}\,\delta_{q,Q},  
\end{equation}
\begin{equation}
  \label{eq:lj38}
  {M}^-_{nn'}(q) = 
  \frac{\left(\prod\limits_{i<j}^{R}\sin^{2}\frac{k_{i}^{n}-k_{j}^{n}}{2}\right)
\left(\prod\limits_{i<j}^{r}\sin^{2}\frac{k_{i}^{n'}-k_{j}^{n'}}{2}\right)}%
  {\prod\limits_{j=1}^{r}N^{2}\prod\limits_{i=1}^{R}\sin^{2}\frac{k_{i}^{n}-k_{j}^{n'}}{2}}\,\delta_{r,R-1}\,\delta_{q,Q},
\end{equation}
\begin{equation}
  \label{eq:32}
  M_{nn'}^z(q) = \frac{u_{nn'}}{N}\,\delta_{r,R}\,\delta_{q,Q},
\end{equation}
where
\begin{equation}
  \label{eq:lj41}
  Q=\sum_{i=1}^{r}k_i^{n'}- \sum_{j=1}^{R}k_j^n~ ~ \mathrm{mod}~2\pi
\end{equation}
is the momentum transfer during the transition, and where $u_{nn'}=(S_T^z)^2$
with $S_T^z=N/2-R$ from (\ref{eq:15}) if the two sets of fermion momenta
(\ref{eq:31}) are identical, $u_{nn'}=1$ if they differ by exactly one element,
and $u_{nn'}=0$ if they differ by more than one element. The energy
transfer is
\begin{equation}
  \label{eq:1}
  \omega_{nn'}=\sum_{i=1}^r\cos k_i^{n'} -\sum_{j=1}^R\cos k_j^{n}.
\end{equation}

The result (\ref{eq:32}) is elementary in the fermion representation. The
product formulas (\ref{eq:lj37}) and (\ref{eq:lj38}) were derived in
Ref.~\cite{BKMW04} from determinantal expressions obtained via algebraic Bethe
ansatz for the planar $XXZ$-model~\cite{BKM03,KMT99,KBI93}.  In the Bethe ansatz
context the $k_i^n$ and $k_i^{n'}$ were magnon momenta, i.e. solutions of the
Bethe ansatz equations. The general solution was shown to produce (i) real,
non-critical magnon momenta, identical to the fermion momenta (\ref{eq:13a}),
(ii) real or complex critical pairs of magnon momenta (with $k_i+k_j=\pi$),
different from any fermion momentum in the set (\ref{eq:13a}). Critical pairs
only occur for even $N$.

The determinantal expressions from which the product formulas (\ref{eq:lj37})
and (\ref{eq:lj38}) were derived contain factors $\cos\frac{1}{2}(k_i+k_j)$ in
the denominator, which are zero for critical pairs. For this reason the
conversion into the product expressions as reported in Ref.~\cite{BKMW04}
proceeded with the restriction that no critical pairs were present.  Since the
product expressions do no longer contain any such vanishing factors, it is very
well possible that their validity is broader in scope than the original
derivation suggested.  On the basis of numerical tests including the sum rule
\begin{equation}
  \label{eq:lj40a}
  \sum_q\left[\sum_{n'}M_{nn'}^+(q)+\sum_{n''}M_{nn''}^-(q)\right]=N\quad \forall n, 
\end{equation}
we are indeed led to conjecture that the product expressions (\ref{eq:lj37}) and
(\ref{eq:lj38}) are universally valid for even or odd $N$ provided we
replace the critical magnon momenta by the corresponding fermion momenta. A
proof of this conjecture will either have to be an extension of the existing
derivation within the framework of the Bethe ansatz to include the critical
magnon momenta or it will have to be an independent derivation within the
fermion representation.

%
\subsection{Transitions from ground state for even $N$}\label{sec:gsexceven}
\begin{figure}[htb]
  \centering
  \includegraphics[width=60mm]{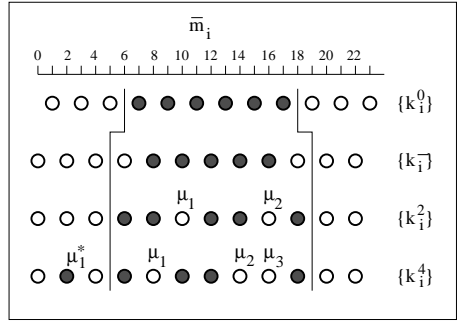}  
  \caption{System with $N=12$ sites. Fermion momenta $k_i=(\pi/N)\bar{m}_i$
    for the lowest energy states at $N_f=N/2$ (row 1) and
    $N_f=N/2-1$ (row 2). Also shown are the fermion momenta for a
    generic 2-spinon state (row 3) and a generic 4-spinon state in the subspace
    with $N_f=N/2-1$. Also shown are segments of the two-pronged fork
  described in the context of Fig.~\ref{fig:fermspinN4}}
  \label{fig:ferm2mspeven}
\end{figure}

%
At $T=0$ only transitions from the lowest energy level survive in expression
(\ref{eq:29}).  For even $N$ the lowest level is the (non-degenerate) spinon
vacuum. Its fermion momentum configuration for $N=12$, now denoted by the set
$\{k_1^0,\ldots,k_{N/2}^0\}$, is illustrated in row 1 of Fig.~\ref{fig:ferm2mspeven}.
From that state the operator $S_q^z$ and $S_q^-$ reach different sets of
excitations. Without loss of generality we assume that $N/4$ is an integer.

In $S_{zz}(q,\omega)_0$ the only non-vanishing transition rates (\ref{eq:32})
pertain to the set of 2-spinon excitations (with $n_+=n_-=1$). In the fermion
representation they are 1-particle-1-hole states relative to the ground state,
generated from row 1 of Fig.~\ref{fig:ferm2mspeven} by moving exactly one
fermion from the inside domain to the outside domain. The total number of such
states is $N^2/4$, each contributing the amount $1/N$ to the spectral weight.
Parenthetically, let us mention the polymer fluctuation operator $P_q^{2M},
M=1,2,3,\ldots$ introduced in Ref.~\cite{DKSM05}.  This operator only allows
transitions from the spinon vacuum to the set of $2M$-spinon excitations (with
$n_+=n_-=M$), i.e. to the set of $M$-particle-$M$-hole excitations in the
fermion representation.

Our main focus here is on $S_{- +}(q,\omega)_0$. Note that we have $S_{- +}(q,\omega)_0
=S_{+ -}(q,\omega)_0$ due to reflection symmetry in spin space, even though the two
functions involve different sets of excited states. All excitations in the
invariant subspace with $N_f=N/2-1$ contribute spectral weight to $S_{-
  +}(q,\omega)_0$. For their systematic generation it is useful to start from the
state with the lowest energy in that subspace. It has spinon composition $n_+=2,
n_-=0$ and its fermion momenta $\{k_i^-\}$ are illustrated in row 2 of
Fig.~\ref{fig:ferm2mspeven}.

The complete set of 2-spinon states in the same invariant subspace can now be
described by two even-valued integer parameters $N/2\leq \mu_1< \mu_2\leq 3N/2$ as
illustrated in row 3 of Fig.~\ref{fig:ferm2mspeven}. They mark the two vacancies
in the array of fermion momenta $\{k_i^2\}$ in the inside domain.  These vacancies
represent spinons with spin up $(n_+=2)$. The spinon momenta are determined by
the rules given in Sec.~\ref{sec:qpr}.

The 4-spinon excitations relevant for $S_{- +}(q,\omega)_0$ have $n_+=3$ and
$n_-=1$. The spin-up spinons are described by three even-valued integer
parameters $N/2\leq \mu_1< \mu_2 < \mu_3\leq 3N/2$ as shown in row 4 of
Fig.~\ref{fig:ferm2mspeven}. The spin-down spinon is described by the
even-valued integer parameter $\mu_1^*$ in the outside domain i.e. with range
$0\leq\mu_1^*<N/2$ or $3N/2<\mu_1^*<2N$. In generalization to this recipe, the
$m$-spinon excitations contributing to $S_{- +}(q,\omega)_0$ have $n_+=m/2+1,
n_-=m/2-1$, generated by moving $m/2-1$ fermions from the inside domain to
the outside domain (with parameters $\mu_1^*,\ldots,\mu_{m/2-1}^*$) and thus leaving
$m/2+1$ vacancies inside (with parameters $\mu_1,\ldots,\mu_{m/2+1}$). 

%
\subsection{Total $m$-spinon intensity for even $N$}\label{sec:tot2mspint}
%

How is the total spectral intensity in the dynamic spin structure factor $S_{-
  +}(q,\omega)$ divided among the $m$-spinon excitations? From a previous study based
on algebraic analysis \cite{BKM98} we know that in the axial regime of the $XXZ$
antiferromagnet most of the spectral weight is carried by the 2-spinon
excitations alone. On the other hand, previous Bethe ansatz studies
\cite{BKM03,BKMW04} yielded evidence that in the planar regime the 2-spinon
contribution approaches zero in the limit $N\to\infty$.

\begin{figure}[tb]
  \centering
  \includegraphics[width=70mm,angle=-90]{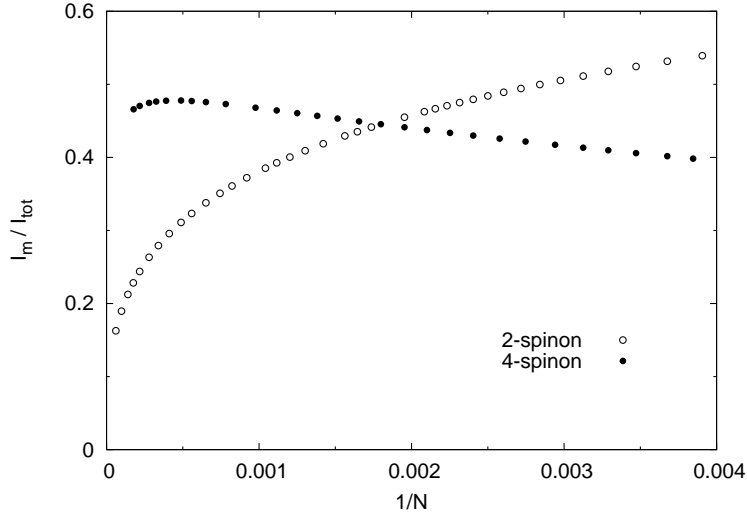}  
  \caption{Relative overall intensity of the 2-spinon excitations ($\circ$) and 4-spinon
    excitations ($\bullet$) in $S_{- +}(q,\omega)_0$  plotted
    versus $1/N$.} 
  \label{fig:totalint}
\end{figure}

By using the transition rate expressions (\ref{eq:lj38}) we have produced
numerical data for the relative 2-spinon intensity $I_2/I_{\rm tot}$ and the
relative 4-spinon intensity $I_4/I_{\rm tot}$, where
\begin{equation}
  \label{eq:14}
  I_{\rm tot}=\sum_q\sum_nM^-_{0n}(q)= \sum_{m=2,4,\ldots}I_m=\frac{N}{2}.
\end{equation}
The data are shown in Fig.~\ref{fig:totalint}. We see that the 2-spinon
intensity accounts for more than half the total intensity in chains with up to
at least $N\simeq300$ sites.  Nevertheless, the data are consistent with the
conclusion that the relative 2-spinon intensity drops to zero in the limit
$N\to\infty$.

The relative contribution of the 4-spinon excitations first rises with $N$. It
levels off at $\sim48\%$ for $N\simeq2000$ and then starts to decrease. In all
likelihood, the relative 4-spinon intensity and, for that matter, any relative
$m$-spinon intensity taken individually, will vanish in the limit $N\to\infty$.
However, the combined 2-spinon and 4-spinon relative intensity stays above 80\%
for chains with up to $N\simeq2000$ sites.

%
\subsection{Asymptotic transition rates for even
  $N$}\label{sec:2mspindsf}  
How is the $m$-spinon intensity of $S_{- +}(q,\omega)_0$ distributed in the
$(q,\omega)$-plane for large $N$? To answer this question we must determine the
explicit dependence of the transition rate expression (\ref{eq:lj38}) on the
spectral parameters $q$ and $\omega$. The product nature of (\ref{eq:lj38}) suggests the
following factorization of the transition rate expression between the spinon
vacuum and an $m$-spinon state (for $m=2,4,\ldots$):
\begin{equation}
  \label{eq:separation}
  M^-\big(\{k_i^{0}\},\{k_i^m\}\big)
  = 
  \underset{\mathrm{scaled \; transition \; rate}}{\underbrace{
      \frac{
        M^-\big(\{k_i^{0}\},\{k_i^m\}\big)
      }{ 
        M^-\big(\{k_i^0\},\{k_i^-\}\big) 
      }
    }}
  \times 
  \underset{\mathrm{scaling \; factor}}{\underbrace{
      \rule[-4mm]{0mm}{9mm}
      M^-\big(\{k_i^0\},\{k_i^-\}\big).
    }}
\end{equation}
Here $\{k_i^0\}$ and $\{k_i^-\}$ are the fermion momenta of the lowest state in
the invariant Hilbert subspaces with $N/2$ fermions and $N/2-1$ fermions,
respectively. The fermion momenta $\{k_i^{m}\}$ of an arbitrary $m$-spinon
excitation differ from the set $\{k_i^-\}$ as explained in the context of
Fig.~\ref{fig:ferm2mspeven}.  In the representation (\ref{eq:separation}) we
use the transition rate between the two reference states (second
factor) as a common scaling factor for all $m$-spinon excitations. In the
scaled transition rates (first factor), huge cancellations take place
because different $m$-spinon excitations differ by no more than $m$ fermion
momenta.

The asymptotic scaling factor must be calculated by expansions of
the product expression
\begin{equation}
  \label{eq:5}
 M^-\big( \big\{k_i^0\big\}, \big\{k_i^-\big\} \big) = 
\sqrt{N} C_{N}(N/2),
\end{equation}
\begin{equation}
  \label{eq:38}
  C_{N}(n) \doteq \prod\limits_{l=1}^{n}
  \frac{\sin^{4l-3}\eta(l-1/2)}{\sin^{4l-1} \eta l },\qquad \eta \doteq \frac{\pi}{N}
\end{equation}
as inferred from (\ref{eq:lj38}) for the two specific states.
The result is of the form
\begin{equation}
  \label{eq:17}
M^-\big( \big\{k_i^0\big\}, \big\{k_i^-\big\} \big) \stackrel{a}{=} \sqrt{N}C,  
\end{equation}
where here and henceforth the symbol ``$\stackrel{a}{=}$" denotes an
asymptotic equality for large $N$ that ignores corrections of the kind
$\times[1+\mathcal{O}(N^{-1})]$. The constant 
\begin{eqnarray}
  \label{eq:18}
  C & \doteq & \lim_{N\to\infty} C_{N}(N/2)
  \\ \nonumber & = & \sqrt{\pi}\exp\left( \frac{\ln2}{6}+6\zeta^\prime(-1) \right)
  =  0.73739071\ldots,
\end{eqnarray}
where $\zeta(z)$ is the Riemann zeta function is familiar from previous
calculations of correlation functions for the $XX$ model by a different approach
\cite{MBA71}. We now present explicit results for the 2-spinon and 4-spinon
transition rates.

%
\subsection{Asymptotic $2$-spinon transition rates}\label{sec:2spp}
%
The scaled transition rate for a 2-spinon excitation can be reduced to the
following product:
\begin{eqnarray}
  \label{eq:th140}
  \frac{
    M^-\big(\{k_i^0\},\{k_i^2\}\big)
  }{
    M^-\big(\{k_i^0\},\{k_i^-\}\big)
  } 
  = N^{2}\sin^{2}\frac{\kappa_{1}-\kappa_{2}}{2}
  \prod_{j=1}^{2} \phi(\psi_{j})\phi(N/2-\psi_{j})
\end{eqnarray}
with the function
\begin{equation}
  \label{eq:37}
  \phi(n) \doteq \prod_{l=1}^{n} \frac{\sin^{2}\eta(l-\frac{1}{2})}{\sin^{2}\eta l},
  \qquad n\in \mathbb{N}^{+},
  \qquad 
  \phi(0) \doteq 1,
\end{equation}
depending on two parameters $\mu_1$ and
$\mu_2$ via
\begin{equation}
  \label{eq:th145}
  \psi_i = \frac{\mu_i}{2}-\frac{N}{4},\quad \kappa_i=\eta \mu_i, \qquad i=1,2.
\end{equation}
Performing the limit $N\to\infty$
is delicate because of singularities along the spectral boundaries of the
2-spinon excitations. To leading order in $N$, expression (\ref{eq:th140})
becomes
\begin{equation}
  \label{eq:19}
   \frac{M^-\big(\{k_i^0\},\{k_i^2\}\big)}{M^-\big(\{k_i^0\},\{k_i^-\}\big)} \stackrel{a}{=}
\frac{4}{N^2}\frac{\displaystyle \sin^2\frac{\kappa_1-\kappa_2}{2}}{\cos\kappa_1\cos\kappa_2},
\end{equation}
where the two parameters $\kappa_1,\kappa_2$ are related to the energy transfer $\omega$ and
momentum transfer $q$ as follows:
\begin{eqnarray}
  \label{eq:20}
 \omega \stackrel{a}{=} -\sum_{i=1}^{2}\cos\kappa_i, \qquad 
 q = \pi-\sum_{i=1}^{2}\kappa_i ~\mathrm{mod}~2\pi.
\end{eqnarray}
Note that the limits $\{k_i^2\}\to\{k_i^-\}$ and $N\to\infty$ are not
interchangeable. Hence we cannot recover the unit value of the scaled transition
rate by taking the limit $\kappa_1-\kappa_2\to\pi$ and $\kappa_1+\kappa_2\to 2\pi$ in the asymptotic
expression (\ref{eq:19}). 

Assembling the results (\ref{eq:17}) and (\ref{eq:19}) with (\ref{eq:20}) into
(\ref{eq:separation}) yields the 2-spinon transition rate function
\begin{equation}
  \label{eq:21}
M_2^-(q,\omega) \stackrel{a}{=}\frac{4C}{N^{3/2}}\,\frac{4\sin^2(q/2)-\omega^2}{\omega^2-\sin^2q}
\end{equation}
over the (asymptotic) range $|\sin q|\leq\omega\leq2|\sin(q/2)|$ of the 2-spinon spectrum.
The 2-spinon part of the dynamic spin structure factor $S_{- +}(q,\omega)_0$ is then
obtained by multiplying the 2-spinon transition rate function (\ref{eq:21})
with the 2-spinon density of states,
\begin{equation}
  \label{eq:22}
D_2(q,\omega) \stackrel{a}{=}\frac{N}{2\pi}\frac{1}{\sqrt{4\sin^2(q/2)-\omega^2}},  
\end{equation}
yielding
\begin{equation}
  \label{eq:23}
  S_{- +}^{(2)}(q,\omega)_0 \stackrel{a}{=} \frac{2C}{\pi\sqrt{N}}\frac{\sqrt{4\sin^2(q/2)-\omega^2}}{\omega^2-\sin^2q}.
\end{equation}
The interpretation of the asymptotic result (\ref{eq:23}) requires caution. It
is indicative of the 2-spinon spectral-weight distribution in the following
sense: in a histogram plot of $S_{- +}^{(2)}(q,\omega)_0$ based on the transition
rates (\ref{eq:lj38}), the shape of (\ref{eq:23}) comes into better and better
focus as $N$ grows larger, while the relative intensity of this structure fades
away.

%
\subsection{Asymptotic $4$-spinon transition rates}\label{sec:4spp}
%
We start again from the factorized expression (\ref{eq:separation}). The
asymptotic scaling factor remains the same: Eq.~(\ref{eq:17}). The
surviving factors in the scaled transition rate after the massive cancellations
now depend on four parameters (see
Fig.~\ref{fig:ferm2mspeven}):
\begin{eqnarray}
  \frac{M^-\big(\{k_i^0\},\{k_i^4\}\big)}{M^-\big(\{k_i^0\},\{k_i^-\}\big)}
  & = & N^2 \sin^2\left( \frac{\kappa_1^*}{2} -\frac{\pi}{4} \right)
      \frac{\displaystyle \prod_{i<j}^{3}\sin^2 \frac{\kappa_i-\kappa_j}{2}}%
      {\displaystyle \prod_{i=1}^{3}\sin^2 \frac{\kappa_1^*-\kappa_i}{2}}
  \nonumber \\
  & &\hspace*{-1.0cm}\times 
  \frac{\phi(N/4-\mu_{1}^{*}/2)}{\phi(3N/4-\mu_{1}^{*}/2)}
  \prod_{j=1}^3 \phi(\psi_{j})\phi(N/2-\psi_{j})
\end{eqnarray}
where 
\begin{equation}
  \label{eq:th128}
  \psi_i = \frac{\mu_i}{2}-\frac{N}{4},\quad \kappa_i=\eta \mu_i, \quad i=1,2,3,
  \qquad
  \kappa_1^*=\eta \mu_1^*.
\end{equation}
From this result we have derived the following asymptotic
expression for the scaled 4-spinon transition rates:
\begin{eqnarray}
  \label{eq:th120}
  \hspace*{-5mm} \frac{M^-\big(\{k_i^0\},\{k_i^4\}\big)}{M^-\big(\{k_i^0\},\{k_i^-\}\big)}
   &\stackrel{a}{=}&
  \frac{4}{N^4}\,
\frac{ \cos(\kappa_1^*) }{\displaystyle\prod_{i=1}^{3}\cos(\kappa_i) } 
      \frac{\displaystyle \prod_{i<j}^{3}\sin^2 \frac{\kappa_i-\kappa_j}{2}}%
      {\displaystyle \prod_{i=1}^{3}\sin^2 \frac{\kappa_1^*-\kappa_i}{2}}
\end{eqnarray}
The asymptotic dependence of the energy and momentum transfer of 4-spinon
excitations on the four parameters is
\begin{eqnarray}
  \label{eq:th117}
  \omega \stackrel{a}{=} \cos\kappa_1^* - \sum_{i=1}^{3}\cos\kappa_i,
\qquad 
  q =  \pi+\kappa_1^*-\sum_{i=1}^{3}\kappa_i ~ \mathrm{mod}~ 2\pi.
\end{eqnarray}
From inspection of the 2-spinon result (\ref{eq:19}) and the 4-spinon result
(\ref{eq:th120}) we can already see features of the general structure for the
$m$-spinon result emerging.

\begin{figure}[htb]
  \centering
  \includegraphics[width=70mm,angle=-90]{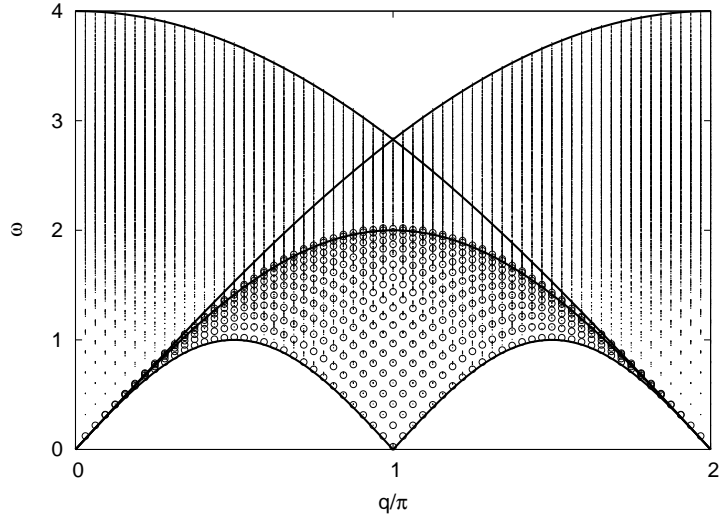}  
  \caption{Excitation energy versus wave number of all 2-spinon $(\circ)$ and
    4-spinon $(\cdot)$ excitations for $N=64$. The solid lines are the spectral
    boundaries for $N\to\infty$.}
  \label{fig:spectrum2p4}
\end{figure}
The range of the 4-spinon excitations in the $(q,\omega)$-plane
in relation to the range of the 2-spinon excitations is illustrated in
Fig.~\ref{fig:spectrum2p4} for $N=64$ and asymptotically for $N\to\infty$. We can
discern the nonuniform density of 2-spinon states (\ref{eq:22})
in the distribution of circles. Furthermore, from the distribution of dots we
see that the density of 4-spinon excitations is very small near the spectral
threshold and remains relatively small at energies $\omega\lesssim 1$.
In the limit $N\to\infty$, the 2-spinon and 4-spinon excitations have a common
spectral threshold,
\begin{equation}
  \label{eq:2}
  \epsilon_{2L}(q)=\epsilon_{4L}(q)=|\sin q|,
\end{equation}
and different upper boundaries,
\begin{equation}
  \label{eq:34}
  \epsilon_{2U}(q)=2\left|\sin\,\frac{q}{2}\right|,
\end{equation}
\begin{equation}
  \label{eq:4}
  \epsilon_{4U}(q)=4\,\mathrm{max}\left[\;\left|\sin\,\frac{q}{4}\right|, 
\left|\sin\,\frac{q-2\pi}{4}\right|\;\right].
\end{equation}

%
\subsection{Transitions from the ground state for odd
  $N$}\label{sec:spin1c3dsf}  
%
For odd $N$ the ground state is fourfold degenerate. Two of its vectors
(labeled $A$ and $A'$) are located in the invariant subspace with
$N_f=(N-1)/2$ fermions and the other two vectors (labeled $B$ and
$B'$) in the subspace with $N_f=(N+1)/2$ fermions.  We then have to
consider transitions from each of the four ground-state vectors to the sets of
$m$-spinon excitations with $m=1,3,\ldots$ in the accessible invariant subspaces.
Without loss of generality we assume that $(N+1)/4$ is an integer.

\begin{figure}[htb]
  \centering
  \includegraphics[width=62mm]{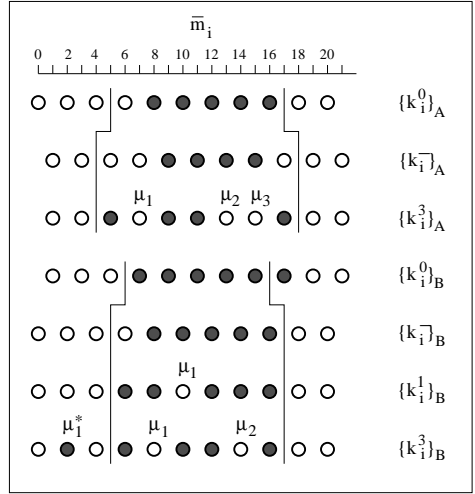}  
  \caption{$XX$ chain with $N=11$ sites. Fermion momenta $k_i=(\pi/N)\bar{m}_i$
    for one of two lowest energy states at $N_f=(N-1)/2$ (rows 1 and 5),
    $N_f=(N-3)/2$ (row 2), and $N_f=(N+1)/2$ (row 4).  The other state in each
    case (henceforth referred to by subscripts $A'$ or $B'$) is obtained via
    reflection at the line $\bar{m}_i=11$. Also shown are the fermion momenta
    for a generic 1-spinon state in the subspace with $N_f=(N-1)/2$ (row 6), a
    generic 3-spinon state in the subspace with $N_f=(N-3)/2$ (row 3), and a
    generic 3-spinon state in the subspace with $N_f=(N-1)/2$ (row 7).}
  \label{fig:fermmspodd}
\end{figure}

The fermion momentum configurations of two of the four ground states are shown
in rows 1 and 4 of Fig.~\ref{fig:fermmspodd} for $N=11$. It will suffice to
consider these two vectors (with fermion momenta $\{k_1^0,\ldots,k_{N/2-1}^0\}_A$ and
$\{k_1^0,\ldots,k_{N/2+1}^0\}_B$) and calculate from them the functions $S_{\mu\nu}(q,\omega)_A$
and $S_{\mu\nu}(q,\omega)_B$. Symmetry dictates that the other two ground state vectors
yield the functions $S_{\mu\nu}(q,\omega)_{A'}=S_{\mu\nu}(-q,\omega)_A$ and
$S_{\mu\nu}(q,\omega)_{B'}=S_{\mu\nu}(-q,\omega)_B$. Hence the $T=0$ dynamic spin structure factor
(\ref{eq:29}) for odd $N$ can be rewritten as
\begin{equation}
  \label{eq:24}
  \hspace*{-17mm} S_{\mu\nu}(q,\omega) = \frac{1}{4}\left[ S_{\mu\nu}(q,\omega)_A + S_{\mu\nu}(-q,\omega)_A + S_{\mu\nu}(q,\omega)_B
    + S_{\mu\nu}(-q,\omega)_B\right].
\end{equation}
Different sets of excitations are again relevant for $\mu\nu=zz,+ -, - +$.

In $S_{zz}(q,\omega)_A$ the non-vanishing transition rates (\ref{eq:32}) now include
all $(N+1)/2$ 1-spinon states (with $n_+=1$), including the ground-state vector
$A$ itself, and a subset of $(N-1)^2/4$ 3-spinon states (with $n_+=2,n_-=1$),
all in the same invariant subspace. In row 1 of Fig.~\ref{fig:fermmspodd} the
1-spinon states are obtained by moving the vacancy to any site in the inside
domain, whereas the contributing 3-spinon excitations are obtained by leaving
the vacancy inside as is and moving one particle from anywhere inside to
anywhere outside. The relevant spectrum of $S_{zz}(q,\omega)_B$ is identified in like
manner by starting from row 4 and interchanging the roles of the two domains and
the roles of particles and vacancies.

As in Sec.~\ref{sec:gsexceven} our main focus is on $\mu\nu=- +$. The spectrum for
$S_{- +}(q,\omega)_A$ includes all eigenstates with $N_f=(N-3)/2$ and the spectrum
for $S_{- +}(q,\omega)_B$ all eigenstates with $N_f=(N-1)/2$.  For a systematic
generation of all relevant excitations we again consider one state each in the
invariant subspaces with $N_f=(N-3)/2$ and $N_f=(N-1)/2$ as reference states for
the scaled matrix elements. These two states have fermion momentum
configurations $\{k_1^-,\ldots,k_{(N-3)/2}^-\}_A$ and $\{k_1^-,\ldots,k_{(N-1)/2}^-\}_B$ as
illustrated (for $N=11$) in rows 2 and 5, respectively, of
Fig.~\ref{fig:fermmspodd}.

Reference state $\{k_i^-\}_A$ is a 3-spinon state (with
$n_+=2,n_-=1$) whereas reference state $\{k_i^-\}_B$ is a 1-spinon state (with
$n_+=1$).  The remaining states in the subspace of $\{k_i^-\}_A$ are $m$-spinon
states for $m=3,5,\ldots$ with $n_+=(m+3)/2, n_-=(m-3)/2$, whereas the remaining
states in the subspace of $\{k_i^-\}_B$ are $m$-spinon states for $m=1,3,\ldots$
with $n_+=(m+1)/2, n_-=(m-1)/2$. The 1-spinon and 3-spinon states thus
generated are illustrated for $N=11$ in rows 3, 6, and 7 of
Fig.~\ref{fig:fermmspodd}. Note in particular that 1-spinon excitations only
occur in $S_{- +}(q,\omega)_B$ but not in $S_{- +}(q,\omega)_A$.

%
%
We have produced extensive finite-$N$ data for the relative 1-spinon intensity
$I_1^B/I_{\rm tot}^B$ and the relative 3-spinon intensities $I_3^A/I_{\rm tot}^A$
and $I_3^B/I_{\rm tot}^B$, where
\begin{eqnarray}
  \label{eq:25}
    I_{\rm tot}^A &=& \sum_q\sum_nM^-_{An}(q)= \sum_{m=3,5,\ldots}I_m^A=\frac{N-1}{2}, \nonumber \\
    I_{\rm tot}^B &=& \sum_q\sum_nM^-_{Bn}(q)= \sum_{m=1,3,\ldots}I_m^B=\frac{N+1}{2}. 
\end{eqnarray}

\begin{figure}[bt]
  \centering
  \includegraphics[width=70mm,angle=-90]{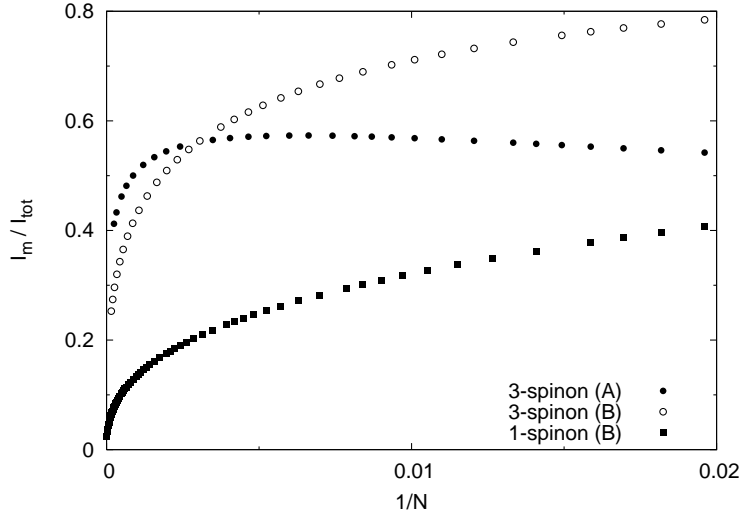}  
  \caption{Relative overall intensity of the 1-spinon
    excitations $(\blacksquare$) and 3-spinon excitations $(\circ)$ in $S_{- +}(q,\omega)_B$ , and
    of the 3-spinon excitations $(\bullet$) in $S_{- +}(q,\omega)_A$, all plotted versus
    $1/N$.}
  \label{fig:totalintodd}
\end{figure}

The data are shown in Fig.~\ref{fig:totalintodd}. We see that the contribution
of the 1-spinon excitations to the total intensity remains significant for
chains with up to several hundred sites and the contribution of 3-spinon
excitations for chains up to several thousand sites. Nevertheless, the data give
a clear indication that both contributions vanish in the limit $N\to\infty$.

%
\subsection{Asymptotic transition rates for odd
  $N$}\label{sec:astrodd}  
In the factorization (\ref{eq:separation}) we now have to
distinguish two scaling factors for the degenerate ground state. Both must be
calculated anew, from the product expressions
\begin{equation}
  \label{eq:6}
  M^-\big(\{k_i^0\}_A,\{k_i^-\}_A\big) =  
  \sqrt{N} C_{N}((N-1)/2)\cos^2 \frac{\eta}{2}, 
\end{equation}
\begin{equation}
  \label{eq:7}
M^-\big(\{k_i^0\}_B,\{k_i^-\}_B\big) =
\sqrt{N} C_{N}((N-1)/2),
\end{equation}
with $C_{N}((N-1)/2)$ defined in (\ref{eq:38}). The asymptotic results are again of the form
\begin{equation}
  \label{eq:8}
  M^-\big(\{k_i^0\}_A,\{k_i^-\}_A\big) \stackrel{a}{=} 
M^-\big(\{k_i^0\}_B,\{k_i^-\}_B\big)\stackrel{a}{=}\sqrt{N}C
\end{equation}
with the constant $C$ from (\ref{eq:18}).

The scaled transition rate for a 1-spinon excitation is reduced to
\begin{eqnarray}
  \label{eq:166}
  \frac{
    M^-\big( \{k_i^0\}_B, \{k_i^1\}_B \big)
  }{
    M^-\big( \{k_i^0\}_B, \{k_i^-\}_B \big)
  } 
  = N\cos^2(\eta \psi_1) \phi(\psi_{1})\phi((N-1)/2-\psi_{1}),
\end{eqnarray}
where 
\begin{equation}
  \label{eq:9}
  \psi_1=\frac{\mu_1}{2}-\frac{N+1}{4},\quad \kappa_1=\eta \mu_1.
\end{equation}
Asymptotically for $N\to\infty$ Eq.~(\ref{eq:166}) becomes
\begin{equation}
  \label{eq:10}
   \frac{M^-\big( \{k_i^0\}_B, \{k_i^1\}_B \big)}{M^-\big( \{k_i^0\}_B,
     \{k_i^-\}_B \big)} \stackrel{a}{=} \frac{1}{N}\cot\left(\frac{\kappa_1}{2}-\frac{\pi}{4}\right).
\end{equation}
The 1-spinon energy and momentum transfers are
\begin{equation}
  \label{eq:11}
  \omega \stackrel{a}{=}-\cos\kappa_1,\qquad q \stackrel{a}{=}\frac{3\pi}{2}-\kappa_1.
\end{equation}
Note again that the limits $\{k_i^1\}_B\to\{k_i^-\}_B$ and $N\to\infty$ are no longer
interchangeable in the asymptotic result.  The 1-spinon transition rate
function thus becomes
\begin{equation}
  \label{eq:12}
  M_1^-(q,\omega)_B \stackrel{a}{=}\frac{C}{\sqrt{N}}\tan\frac{q}{2}.
\end{equation}
Taking into account the fourfold ground-state degeneracy, we can write the asymptotic
expression for the 1-spinon part of the dynamic spin structure factor in the
form
\begin{equation}
  \label{eq:13}
  S_{- +}^{(1)}(q,\omega)\stackrel{a}{=} \frac{C}{4\sqrt{N}}\tan\frac{q}{2}\delta\left(\omega-|\sin q|\right).
\end{equation}

Different sets of 3-spinon states are reached via $S_q^-$ from the ground-state
components $A$ and $B$. Hence we must work with two different results for the
scaled transition rates. For ground-state $B$ we obtain the expression
\begin{eqnarray}
  \label{eq:226}
 && \hspace*{-15mm}\frac{
    M^-\big(\{k_i^0\}_B,\{k_i^3\}_B\big)  
  }{
    M^-\big(\{k_i^0\}_B,\{k_i^-\}_B\big)  
  } 
   = 
  N\tan^{2}(\eta \psi_{1}^{*})\sin^{2}(\eta(\psi_{1}-\psi_{2})) 
  \prod_{j=1}^{2}\frac{\cos^2(\eta\psi_{j})}{\sin^{2}(\eta(\psi_{1}^{*}-\psi_{j}))}
  \nonumber \\
  && \times 
  \frac{\phi(-\psi_{1}^{*})}{\phi((N-1)/2-\psi_1^*)}
  \prod_{j=1}^{2}\phi(\psi_{j})\phi((N-1)/2-\psi_{j})
\end{eqnarray}
depending on the three parameters
\begin{equation}
  \label{eq:16}
  \psi_1=\frac{\mu_1}{2}-\frac{N+1}{4},\quad
  \psi_2=\frac{\mu_2}{2}-\frac{N+1}{4},\quad
\psi_1^*=\frac{\mu_1^*}{2}-\frac{N+1}{4}.\quad
\end{equation}
The asymptotic expansion of (\ref{eq:226}) yields the following leading term,
now parameterized by $\kappa_1=\eta \mu_1$, $\kappa_2=\eta \mu_2$, and $\kappa_1^*= \eta \mu_1^*$:
\begin{eqnarray}
  \label{eq:210}
  \hspace*{-1.0cm}\frac{
    M^-\big(\{k_i^0\}_B,\{k_i^3\}_B\big)
  }{
    M^-\big(\{k_i^0\}_B,\{k_i^-\}_B\big)
  }
  &\stackrel{a}{=}&
  \frac{1}{N^3}\,
  \frac{\displaystyle
    \sin^2\big(\frac{\kappa_1-\kappa_2}{2}\big)
  }{\displaystyle \prod_{i=1}^{2}
    \tan\big(\frac{\kappa_i}{2}-\frac{\pi}{4}\big)
  } 
  \frac{\displaystyle 
    \tan\big(\frac{\pi}{4}-\frac{\kappa^*_1}{2}\big)
  }{\displaystyle \prod_{i=1}^{2}
    \sin^2\big(\frac{\kappa^*_1-\kappa_i}{2}\big)
  }.
\end{eqnarray}
The energy transfer and momentum transfer in the asymptotic regime are
\begin{eqnarray}
  \label{eq:213}
  \omega  \stackrel{a}{=}  \cos\kappa^*_1-\sum_{i=1}^{2}\cos\kappa_i,
  \qquad 
  q  \stackrel{a}{=}  { \kappa^*_1-\sum_{i=1}^{2}\kappa_i-\frac{\pi}{2} }\quad \mathrm{mod}~2\pi.
\end{eqnarray}

\begin{figure}[htb]
  \centering
  \includegraphics[width=70mm,angle=-90]{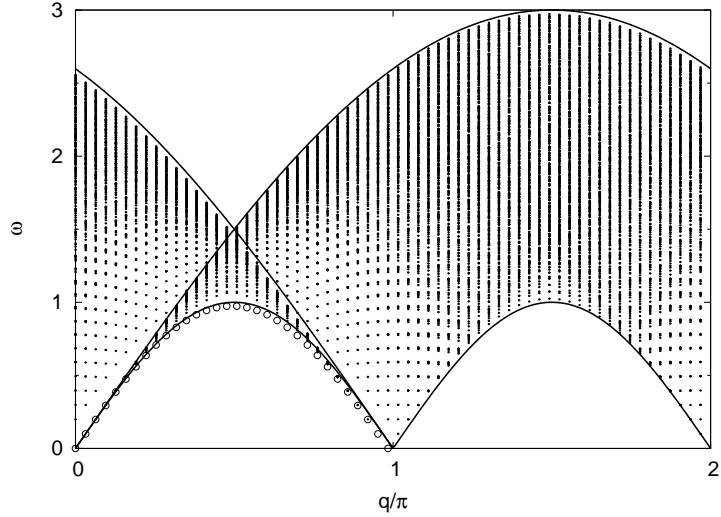}  
  \caption{Excitation energy versus wave number of all 1-spinon
    excitations ($\circ$) and 3-spinon excitations ($\cdot$) from ground-state $B$ for
    $N=63$. The solid lines are the 3-spinon continuum boundaries for
    $N\to\infty$. The corresponding 1-spinon and 3-spinon spectrum from ground-state
  $B'$ is the mirror image reflected at $q/ \pi=1$ of the spectrum shown.}
  \label{fig:1and3spinonB}
\end{figure}

The range in the $(q,\omega)$-plane of the 1-spinon and 3-spinon excitations from
ground-state $B$ is shown in Fig.~\ref{fig:1and3spinonB} for $N=63$. The
corresponding spectrum reached from ground-state $B'$ is the one with
wave numbers $q$ replaced by $2\pi-q$.  In the limit $N\to\infty$ the single branch
of 1-spinon excitations coincides with the lower boundary of the 3-spinon
continuum,
\begin{equation}
  \label{eq:33}
  \epsilon_1(q)=\epsilon_{3L}(q)=|\sin\,q|.
\end{equation}
The upper boundary of the 3-spinon continuum is
\begin{equation}
  \label{eq:26}
 \epsilon_{3U}(q)=3\,\mathrm{max}\left[\;\left|\sin\,\frac{q}{3}\right|, 
\left|\sin\,\frac{q-2\pi}{3}\right|\;\right].
\end{equation}

The scaled transition for 3-spinon transitions from ground-state $A$ reads
\begin{eqnarray}
  \label{eq:103}
   \frac{ M^-\big(\{k_i^3\}_A,\{k_i^0\}_A\big) }
  { M^-\big(\{k_i^-\}_A,\{k_i^0\}_A\big) } &=& 
  N^3\sin^2\frac{\eta}{2}\,
  \frac{\displaystyle \prod_{i<j}^{3}\sin^{2}(\eta(\psi_{i}-\psi_{j}))}%
  {\displaystyle \prod_{j=1}^{3}\sin^{2}(\eta(\psi_{j}-1/2))} \nonumber \\
 &\times&  \prod_{j=1}^{3}\phi(\psi_{j})\phi((N+1)/2-\psi_{j})
\end{eqnarray}
depending on the three parameters
\begin{equation}
  \label{eq:35}
  \psi_i= \frac{\mu_i}{2}-\frac{N-1}{4},\quad i=1,2,3.
\end{equation}
The leading term of the asymptotic expansion becomes a function of
the three variables $\kappa_i=\eta \mu_i$:
\begin{eqnarray}
  \label{eq:109}
  && \hspace*{-15mm}
  \frac{ M^-\big(\{k_i^3\}_A,\{k_i^0\}_A\big) }
  { M^-\big(\{k_i^-\}_A,\{k_i^0\}_A\big) } \stackrel{a}{=}  
  \frac{\pi^2}{4N^5}
  \frac{\displaystyle \prod_{i<j}^{3} \sin^{2}\frac{\kappa_{i}-\kappa_{j}}{2}}%
  {\displaystyle \prod_{i=1}^{3} \cos\left(\frac{\kappa_{i}}{2}
-\frac{\pi}{4}\right)\sin^3\left(\frac{\kappa_{i}}{2}-\frac{\pi}{4}\right)}
\end{eqnarray}
with associated energy and momentum transfer
\begin{eqnarray}
  \label{eq:110}
  \omega  \stackrel{a}{=}  
  -\sum_{i=1}^{3}\cos \kappa_i,
  \qquad 
  q  \stackrel{a}{=} \frac{3\pi}{2}-\sum_{i=1}^{3}\kappa_i\quad \mathrm{mod}~2\pi.
\end{eqnarray}

\begin{figure}[htb]
  \centering
  \includegraphics[width=70mm,angle=-90]{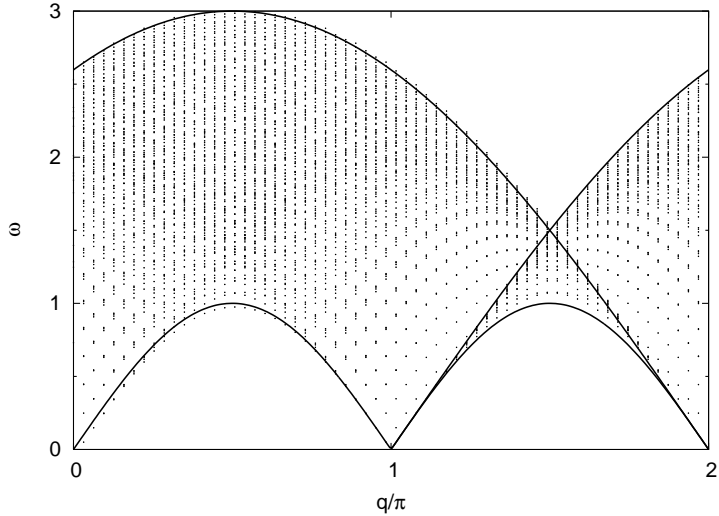}  
  \caption{Excitation energy versus wave number of all 3-spinon excitations
    ($\cdot$) from ground-state $A$ for $N=63$. The solid lines are the 3-spinon
    continuum boundaries for $N\to\infty$. The corresponding 3-spinon spectrum from
    the ground-state component $A'$ is the mirror image reflected at $q/ \pi=1$
    of of the spectrum shown.}
  \label{fig:3spinonA}
\end{figure}

The range in the $(q,\omega)$-plane of the 3-spinon excitations from
ground-state $A$ is shown in Fig.~\ref{fig:3spinonA} for $N=63$. The
corresponding spectrum reached from ground-state component $A'$ is the one with
wave numbers $q$ replaced by $2\pi-q$. In the limit $N\to\infty$ the 3-spinon
continuum boundaries for states $A$ and $A'$ combined are the same as those of
states $B$ and $B'$ combined. 


%
\section{Conclusion and outlook}\label{sec:specwedi}
%

The two main results on the $XX$ model reported in this paper are (i) the exact
one-to-one mapping between fermion and spinon compositions of all eigenstates
and (ii) the applications of exact product expressions for the transition rates
as needed for the calculation of dynamic spin structure factors. The mapping
has provided us with a framework that allows us to explore any property of
interest by using the computationally convenient (free) fermion quasiparticles
and interpret the results in terms of the physically relevant (interacting)
spinon quasiparticles. The very product nature of the transition rate
expressions for the $XX$ model has put us at a huge advantage in the face of
the equivalent determinantal expressions known for the $XXZ$ model in general,
when it comes to the evaluation of explicit numerical results or the
calculation of asymptotic analytic results for large systems.

The applications worked out in Sec.~\ref{sec:dyspistruf} call for a new
assessment of some issues and set the stage for the pursuit of new questions of
interest.  Chief among the issues that need to be reevaluated in the light of
the results reported here concerns the role in general and the observability in
particular of specific sets of spinon quasiparticles in dynamic spin structure
factors and related quantities. 

The evidence presented in Sec.~\ref{sec:dyspistruf} suggests that the relative
intensity of any individual $m$-spinon contribution to the dynamic spin
structure factor $S_{- +}(q,\omega)$ at $T=0$ vanishes in the limit $N\to\infty$. This
conclusion, which is expected to hold true not only for the $XX$ model but
within the entire planar regime of the $XXZ$ model \cite{BKM03}, is in contrast
to what is known on rigorous grounds for the $XXZ$ model in the axial regime
including the isotropic limit ($XXX$ model): the relative intensity of the
2-spinon excitations alone amounts to 73\% in the Heisenberg limit and
gradually increases to 100\% in the Ising limit \cite{KMB+97,BKM98}.

The dominant singularities in the dynamic structure factor have been shown to
be extractable from the 2-spinon part alone in the $XXX$ model \cite{KMB+97}.
This is manifestly not the case in the $XX$ model and unlikely the case for the
planar $XXZ$ model in general.  A fair amount of exact information about the
singularity structure of $S_{- +}(q,\omega)$ at $T=0$ can be gleaned from existing
results \cite{MBA71,VT78,MS84a,MPS83a,MPS83b}. One important task will be to
assemble the asymptotic $m$-spinon results from the approach taken here to
recover known exact results for $N\to\infty$. A hopeful sign for this endeavor is
that the constant $C$ as given in (\ref{eq:18}), which plays a crucial role in
exact results for $N\to\infty$, already emerges from the asymptotic $m$-spinon
expressions worked out here.
  
Dynamic spin structure factors are accessible more or less directly to
experimental investigations via neutron scattering, NMR, ESR, light scattering,
and other probes in the context of quasi-one-dimensional magnetic compounds.
Even small amounts of impurities or other defects will turn a crystalline
sample into an ensembles of mesoscopic chains of various lengths $(N\sim
10^2-10^3)$.  Under what circumstances does the distinction between a physical
ensemble of mesoscopic chains and a statistical ensemble of macroscopic chains
matter?

If the relative intensity of the 2-spinon contribution is dominant for
mesoscopic chains as well as for macroscopic chains, which is the case for the
$XXX$ model and the axial $XXZ$ model, then the distinction has minimal
impact. If, on the other hand, the relative 2-spinon intensity is dominant for
chains of length $N\sim10^2$ and the relative 4-spinon intensity is dominant for
chains of length $N\sim10^3$ but both contributions taken individually or
combined become negligible in macroscopic chains, as is the case in the $XX$
model, then the distinction cannot be ignored. Hence the significance of the
results presented in Sec.~\ref{sec:dyspistruf}.

In conclusion we mention two further unresolved issues to which
the work reported here has led. They are the subject of work currently in
progress. A first set of questions calls for a derivation of the product
expressions (\ref{eq:lj37})-(\ref{eq:lj38}) and for the establishment of
connections to the many exact results for dynamic correlation functions already
known for the $XX$ model. A second set of questions calls for a detailed study of
the spinon interaction and for the exploration of the physical relevant
quasiparticles whose physical vacuum is the $XX$ ground state in a magnetic
field. 

%
\ack
%
Financial support from the DFG Schwerpunkt \textit{Kollektive
  Quantenzust{\"a}nde in elektronischen 1D {\"U}bergangsmetallverbindungen}
(for M.K.) and from the Graduiertenkolleg  \textit{Darstellungstheorie und
  ihre Anwendungen in Mathematik und Physik} (for K.W.)
is gratefully acknowledged.  G.M. thanks Prof. Dr. J. Stolze for useful discussions.

\section*{References}    

\end{document}